\documentclass[sigconf]{acmart}

\usepackage{booktabs} 
\usepackage[utf8]{inputenc} 
\usepackage{listings}
\usepackage{color}
\usepackage{forest}
\usepackage{pifont}
\usepackage{xcolor}

\acmPrice{15.00}

\copyrightyear{2019}
\acmYear{2019}
\setcopyright{acmcopyright}
\acmConference[ETRA 2019]{ETRA, Eye Tracking Research and Applications}{June 2019}{Denver, Colorado (USA)}
\acmDOI{doi}
\acmISBN{978-1-4503-5706-7}

\definecolor{codegreen}{rgb}{0,0.6,0}
\definecolor{codegray}{rgb}{0.5,0.5,0.5}
\definecolor{codepurple}{rgb}{0.58,0,0.82}
\definecolor{backcolour}{rgb}{0.98,0.98,0.98}

\newcommand{\cmark}{\textcolor{green!80!black}{\ding{51}}}
\newcommand{\xmark}{\textcolor{red}{\ding{55}}}

\lstdefinestyle{mystyle}{
    backgroundcolor=\color{backcolour},   
    commentstyle=\color{codegreen},
    keywordstyle=\color{magenta},
    numberstyle=\tiny\color{codegray},
    stringstyle=\color{codepurple},
    basicstyle=\scriptsize\ttfamily,
    breakatwhitespace=false,         
    breaklines=true,                 
    captionpos=b,                    
    keepspaces=true,                 
    numbers=left,                    
    numbersep=5pt,                  
    showspaces=false,                
    showstringspaces=false,
    showtabs=false,                  
    tabsize=2
}

\setcitestyle{square}

\begin{document}

\title[TobiiGlassesPySuite: An open-source suite for using the Tobii Pro Glasses 2]{TobiiGlassesPySuite: An open-source suite for using the Tobii Pro Glasses 2 in eye-tracking studies}

\author{Davide De Tommaso}
\affiliation{%
  \department{Center for Human Technologies}
  \institution{Istituto Italiano di Tecnologia}}
\email{davide.detommaso@iit.it}

\author{Agnieszka Wykowska}
\affiliation{%
  \department{Center for Human Technologies}
  \institution{Istituto Italiano di Tecnologia}}
\email{agnieszka.wykowska@iit.it}

\renewcommand{\shortauthors}{De Tommaso et al.}

\begin{abstract}
In this paper we present the TobiiGlassesPySuite, an open-source suite we implemented for using the Tobii Pro Glasses 2 wearable eye-tracker in custom eye-tracking studies. We provide a platform-independent solution for controlling the device and for managing the recordings. The software consists of Python modules, integrated into a single package, accompanied by sample scripts and recordings. The proposed solution aims at providing additional methods with respect to the manufacturer's software, for allowing the users to exploit more the device's capabilities and the existing software. Our suite is available for download from the repository indicated in the paper and usable according to the terms of the GNU GPL v3.0 license.
\end{abstract}

\ccsdesc[300]{Human-centered computing~Human computer interaction (HCI)}
\ccsdesc[300]{Human-centered computing~User studies}
\ccsdesc[300]{Human-centered computing~Empirical studies in HCI}

\keywords{eye-tracking, human-computer interaction, Tobii Pro Glasses 2, wearable eye-tracker, wearable computing, open-source}

\copyrightyear{2019} 
\acmYear{2019} 
\setcopyright{acmcopyright}
\acmConference[ETRA '19]{2019 Symposium on Eye Tracking Research and Applications}{June 25--28, 2019}{Denver , CO, USA}
\acmBooktitle{2019 Symposium on Eye Tracking Research and Applications (ETRA '19), June 25--28, 2019, Denver , CO, USA}
\acmPrice{15.00}
\acmDOI{10.1145/3314111.3319828}
\acmISBN{978-1-4503-6709-7/19/06}

\maketitle

\section{Introduction}
The recent developments in eye-tracking technologies have produced remarkable progress especially on wearable devices. As a matter of fact, recent market studies predict a notable growing of interest in wearable eye-tracking technology for the next few years. Nowadays, the commercial products available are built to have the appearance of normal eyeglasses, to ensure an easy wearability and a low weight. They are able to track binocular eye-movements up to 200Hz. Tobii AB, Ergoneers GmbH and Pupil Labs are some of the main companies that produce wearable eye-trackers. Although these devices provide similar features, they differ from each others for performances, namely: gaze sampling frequency, tracking technique, communication protocol and programming tools.

In this paper we present an open-source suite developed for Tobii Pro Glasses 2 \cite{tobiiglasses}, the mobile eye-tracker produced by Tobii AB. The TobiiGlassesPySuite offers a unified solution for controlling the device and for processing data in cross-platform environments. It provides the experimenters with simplified methods for accessing the device and the recordings. In fact, the suite integrates the functionalities of the Tobii Pro Glasses 2 API and, at the same time, hides to the users the implementation details. Our work aims at extending the functionalities provided by the manufacturer software by making the device more suitable to be integrated in custom eye-tracking studies.

The paper is divided in four main sections. In Section \ref{sec:1} we present related open-source solutions supporting the eye-tracking research. In Section \ref{sec:2} we highlight the reasons which motivated the software developments presented in the paper. Section \ref{sec:3} contains a more detailed description of the product, in terms of hardware specifications and software tools available. In Section \ref{sec:4} the open-source suite is presented, focusing more on the functionalities provided than on the implementation details. In conclusion, we discuss current limits and future improvements of the proposed solution.

\section{Related Work}\label{sec:1}
Several open-source solutions have been produced over the years to support eye-tracking research and applications. The presence of high-level development tools, accessible also for non-experienced programmers, has enabled researchers to share the results of their efforts for implementing prototypes of new analysis techniques and custom controllers of eye-tracker equipments. The PyGaze software, for example, is an open-source package for creating eye-tracking experiments using Python \cite{Dalmaijer2014}. PyGaze implements the methods for presenting visual and auditory stimulus and for collecting responses using standard and custom input devices. It is compatible with several commercial eye-tracker devices and provides developers with interfaces to implement their custom controllers. In addition, one of the main advantages of PyGaze lies in being able to access all the libraries and frameworks already available for the Python programming language. A similar open-source framework, written in Python, is GazeParser \cite{Sogo2013}. GazeParser, originally developed for Windows OS, offers the possibility to record eye movements using video-based techniques, to create an eye-tracking study using PsychoPy \cite{PEIRCE20078} and VisionEgg \cite{Straw_2008} experimental control libraries and later to extract fixations and saccades. Regarding data analysis tools, many software packages have been released to address the most common issues, such as: for processing of eye movement data to static and dynamic scenes, for detecting and filtering artefacts, for detecting gaze events, for generating AOIs (Areas-Of-Interest) and for visualizing data using heatmaps and gaze plots. Some of these frameworks are available for: MATLAB (GazeAlyze \cite{Berger2012}, EyeMMV \cite{Krassanakis_Filippakopoulou_Nakos_2014}, EALab \cite{Andreu-Perez2016}, SacLab \cite{CERCENELLI201745}), Python (PyGazeAnalyzer \cite{Dalmaijer2014}, \cite{Mardanbegi2017PSOVISA}), R (ETRAN—R \cite{Zhegallo2015ETRANREP}).

\section{Motivation}\label{sec:2}
The wearable solutions have introduced new technical challenges in eye-tracking technology due to the user-centered point of view. Contrary to the static eye-trackers, the gaze coordinates are not referred to a fixed frame of reference (a display screen for example). The lack of a fixed frame presents a significant challenge for the analyst confronted with a highly complex data stream. In mobile eye-trackers, gaze points are registered with respect to the scene camera (placed in front of the glasses) which captures different images depending on the user point of view. Currently, the definition of the AOIs and the relative gaze mapping relies on computer vision algorithms that have still room of improvements, in terms of robustness and easy to use. Some successful attempts have been already proposed in this direction by integrating gaze data with object recognition \cite{Kurzhals_2017, Benjamins:2018, Pfeiffer:2016} and machine learning algorithms \cite{Wolf2018}. The precision/accuracy of the collected gaze data represents another critical point because it varies depending on the calibration procedure and the target distance. In \cite{MacInnes299925}, they compare calibration accuracy and precision among three commercial models of wearable eye-trackers, including the Tobii Pro Glasses 2. Another important open issue is how to determine fixations and saccades using wearable devices, due to the non-static nature of the observed scene. In \cite{Kasneci:2014}, they propose the use of Hidden Markov and Bayesian Mixture models for an online recognition of gaze events. The results show that these probabilistic methods work well in these scenarios thanks to their ability to adapt to the viewing behavior and changes in the scene. In \cite{Munn:2008}, they present an automated process based on I-VT (Velocity-Threshold Identification) gaze filter, applied to different types of scene and eye motions recorded with a mobile eye-tracker. The obtained results show encouraging performances with respect to manual coding. Tobii provides a commercial software for analysis (the Tobii Pro Lab Analyzer \cite{tobiiprolab}) to extract filtered gaze data based using the Tobii I-VT Fixation (Velocity-Threshold Identification Gaze Filter) \cite{Asaro:1976:POT}, although they did not recommend its use with the Tobii Pro Glasses 2.

Considering the expected growth of interest in wearable eye-tracking technology in the next few years and the number of possible application fields, we believe that open-source solutions may facilitate the access to this technology helping significantly the scientific community to  bring out new solutions. In terms of applications, the Tobii Pro Glasses 2 has proved suitable for use in various fields. Recently, it has been employed in research application fields, such as: Cognitive and Social Psychology \cite{Szulewski2015, Ioannidou2017, Arai2017, Rogers2018}, Visual Attention \cite{Rasmus2017, willemse_wykowska_2018}, Clinical Research \cite{Ramey2017, Roderick2018}, Training \cite{SANCHEZFERRER2017668}.

Our work aims at providing open-source tools for experimenters who want to exploit the capabilities of the Tobii Pro Glasses 2 in their research field. Developers can access the source code, modify it and distribute it according to their needs by following the terms and conditions of the GNU GPL v3.0 license. The TobiiGlassesPySuite is written in Python and it cross-platform, so it does not require any commercial software to be used and can be installed in the most common operating systems. The suite provides easy programming tools and examples to facilitate the integration of the device in custom experimental studies. Another aspect that motivated our choices is the fact that the proposed solution can benefit from the many tools already available for Python, including: scientific computing, machine learning,  data analysis, computer vision and so on. Moreover, TobiiGlassesPySuite is compatible with PsychoPy \cite{PEIRCE20078} and integrated in PyGaze \cite{Dalmaijer2014}, extending its native features with others for the development and post-analysis of eye-tracking experiments.

Due to the diffusion of the Tobii Pro Glasses 2 and the lack of any open-source suite for managing the device, we expect a growing interest from the research community in the proposed solution and potentially significant contributions from developers. More technical details can be found in the GitHub repository \cite{tobiiglassespysuite}.

\begin{figure}[h]
 \centering
 \includegraphics[width=\linewidth]{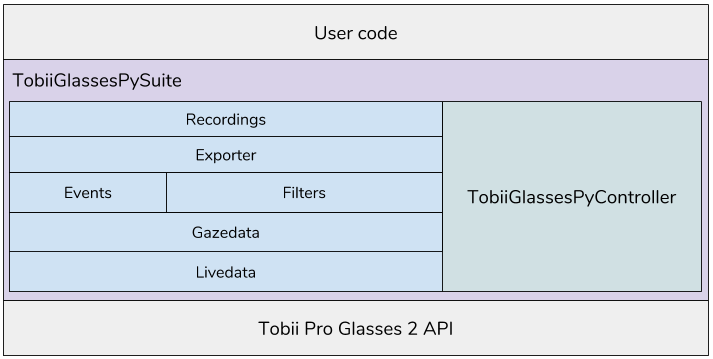}
 \caption{Software architecture of the open-source suite}
 \label{fig:2}
 \end{figure}

\section{Tobii Pro Glasses 2: A general overview}\label{sec:3}
The Tobii Pro Glasses 2 consists of two main units: the \textit{Head Unit}, containing the sensors, and the \textit{Recording Unit}, including an embedded system. The \textit{Head Unit} has the structure of normal eyeglasses and includes a set of infrared projectors and infrared cameras for tracking simultaneously both pupils position, using the two corneal reflection and dark pupil techniques. The \textit{Recording Unit} is an embedded system with connected with the \textit{Head Unit} through a HDMI cable. It is able to send video and data streaming through the network (WiFi or Ethernet) and it stores the recordings in a removable SD card. For a more detailed description of the technical specifications of the product, please refer to the official documentation available at \cite{tobiiglasses}.

The device is also supported by the Tobii Pro Glasses 2 API, a programming interface for accessing all the streamed live data from the Tobii Pro Glasses 2, as well as basic functionalities, such as: managing the calibrations, recordings, projects and participants.

\section{TobiiGlassesPySuite: A Python package for the Tobii Pro Glasses 2}\label{sec:4}
In this section we present the open-source solution we developed for interfacing the Tobii Pro Glasses 2. The source code, available from \cite{tobiiglassespysuite}, it is compatible with Python 2.x and 3.x versions and it is implemented to be platform-independent. Our solution has been tested for working in Windows OS, GNU/Linux and Mac OS systems. The software is embedded in a pip package and available from the Python Package Index (PyPI) repository. This makes the installation relatively simple, requiring only a shell command to complete the operation. The suite consists of two main parts: a stand-alone module for controlling the device (e.g. calibrating, recording, data streaming, etc.), namely the \textit{TobiiGlassesPyController} and a set of other modules for parsing, extracting and filtering the data (\textit{Livedata}, \textit{Gazedata}, \textit{Recordings}, \textit{Filters}, \textit{Exporters} and \textit{Events}). These functionalities are provided by Python classes developed following the OOP (Object Oriented programming) paradigm. All these modules are based and implemented on top of the Tobii Pro Glasses 2 API (see Figure \ref{fig:2}). In such a way, any future change in the official API will not affect the user code, but only the involved classes of the suite. The same API is used by the manufacturer's software for accessing the device and storing the data. This raises us to perform comparisons between our solution and the manufacturer's software because the API operations are managed internally by the device. In the following sections the modules of the suite are presented more in detail.

\begin{center}
\begin{table}
\caption{\label{tab:1} Features provided by the existing software and the proposed solution.}
    \begin{tabular}{ | p{3cm} | p{2cm} | p{2.4cm} |}
    \cline{2-3}
    \multicolumn{1}{c|}{} & \scriptsize \textbf{Tobii Pro Glasses Controller} & \scriptsize \textbf{TobiiGlassesPyController}  \\
    \hline
    \footnotesize GUI & \cmark & \xmark \\ \hline
    \footnotesize Live view with mapped gaze & \cmark & \cmark \\ \hline
    \footnotesize Logged events & \cmark & \cmark \\ \hline
    \footnotesize Tobii Pro Lab custom events & \xmark & \cmark \\ \hline
    \footnotesize Cross-platform & \xmark & \cmark \\ \hline
    \footnotesize Data streaming & \xmark & \cmark \\ \hline
    \end{tabular}
   \end{table}
\end{center}

\subsection{TobiiGlassesPyController}
Tobii provides a free of charge software for controlling the Tobii Pro Glasses 2, namely the \textit{Tobii Pro Glasses Controller}. It allows the users to access a set of features such as: recording, calibrating, and live view using a GUI (Graphical User Interface). The complete list of features can be find in \cite{tobiiglasses}. More experienced users may find some important limitations present in this software that may preclude the use of this device in interactive scenarios or custom eye-tracking experiments. Firstly, the software is compatible only on Windows OS systems. Secondly, the recordings stored in the SD card of the \textit{Recording Unit} are not trivial to retrieve. In fact, data are stored in folders named with unique identifiers generated by the Tobii Pro Glasses 2 API of which the non-expert user is not necessarily aware. Moreover, the recordings are not saved in human-readable formats (CSV or text files), but in the form of JSON objects. Thirdly, gaze data are not accessible on-line during a recording. In order to address these limitations we developed the \textit{TobiiGlassesPyController}, a Python wrapper that uses the functionalities of the Tobii Pro Glasses 2 API, the same way the manufacturer software does, and integrates some additional features. The controller is also available from the repository \cite{tobiiglassespycontroller}. These features, summarized in Table \ref{tab:1}, are shown in the relative examples in Section \ref{sec:4}. In addition to the controller, Tobii provides also a commercial software for analyzing data, the \textit{Tobii Pro Lab Analyzer} \cite{tobiiprolab}. A relevant feature to mention, implemented in the  \textit{TobiiGlassesPyController}, allows to send specific JSON messages during a recording that will be shown in the \textit{Tobii Pro Lab Analyzer} as custom events. This is a feature of particular interest for experimenters who need to analyze separately different conditions in the same recording.

\subsection{Livedata and Gazedata}
As mentioned in the previous section, data processed by the Tobii Pro Glasses 2 are stored in the SD card in the form of JSON objects according to the models described in Tobii Pro Glasses API. They are located in a file named \textit{livedata.json.gz} which contains all the samples processed during a recording. Each sample includes a timestamp (\textit{ts}) and a status flag (\textit{s}) indicating the presence of any anomalies. The \textit{Livedata} module parses the JSON objects, recognizing their type and converting them in specific Python objects. On the other hand, the module \textit{Gazedata} deals with maintaining the gaze data in more efficient structures ordered by timestamps, discarding the not valid or expired samples.

\subsection{Importing and exporting the recordings}
The module \textit{Recordings} provides methods for importing automatically the recordings from the SD card. Recording objects contain information about: project name, participant name, recorded video, gaze positions, gaze directions, head movements, logged events, and so on. The \textit{Exporter} module, instead, implements the mechanisms for exporting the gaze data in CSV format. Specifically, the user can decide to export raw-data or filtered-data. Raw-data consists of a list of ordered samples as they are received by the \textit{TobiiGlassesPyController}, while the filtered-data are the output of one or more filtering functions (such as logged event filters, timestamp filters or fixations/saccades filters). Currently, only the I-DT fixation filter described in \cite{Salvucci:2000} is implemented.

\section{Examples on how to use the suite}\label{sec:5}
In the following, we present Python examples aimed at introducing the user to start using the \textit{TobiiGlassesPySuite} for controlling the eye-tracker and for managing the recordings. These scripts are public available and can be found in \cite{tobiiglassespycontrollerex} and \cite{tobiiglassespysuiteex}. Moreover, in order to explain how to design an eye-tracking study with the \textit{TobiiGlassesPySuite}, a full example of an experiment with recordings is provided.

\begin{minipage}{0.95\linewidth}
\begin{lstlisting}[language=Python, label={lst:1}, caption=A Python example showing some basic functionalities of the TobiiGlassesController, frame=single]
TG = TobiiGlassesPyController()
#TG =TobiiGlassesPyController("192.168.71.50")
#TG =TobiiGlassesPyController("fe80::fffe:ffff:ff00:ff00%eth0")

pjt_name = input("Project's name: ")
pjt_id = TG.create_project(project_name)

ppt_name = input("Participant's name: ")
ppt_id = TG.create_participant(pjt_id, ppt_name)

calib_id = TG.create_calibration(pjt_id, ppt_id)
input("Press ENTER to start calibrating")
TG.start_calibration(calib_id)

res = TG.wait_until_calibration_is_done(calib_id)

if res == False:
	print("Calibration failed!")
	exit(1)

rec_id = tobiiglasses.create_recording(ppt_id)

input("Press ENTER to start recording")

TG.start_streaming()
TG.start_recording(rec_id)

TG.send_logged_event("recording_start")
TG.send_tobiipro_event("recording_event", "start")

while True:
	print("Press 's' to stop the recording, 'g' to get data")
    c = sys.stdin.read(1)
    if c == 's':
    	break
    elif c == 'g':
    	TG.get_data()
    	
TG.send_logged_event("recording_stop")
TG.send_tobiipro_event("recording_event", "stop")

TG.stop_recording(rec_id)
TG.stop_streaming()

\end{lstlisting}
\end{minipage}

\subsection{Connecting through the network (WLAN/LAN)}
The script \textit{connect.py} shows how to connect with the glasses. Using the \textit{TobiiGlassesPyController} constructor without any parameter (see Listing \ref{lst:1}, line 1), a discovery process is started. Specifically, the controller sends periodically discovery packets through all the network interfaces available, until a response from the glasses is captured. In that response the network physical address of the glasses is present. Once the discovery process terminates successfully, the network address of the eye-tracker can be specified as argument of the \textit{TobiiGlassesController} constructor to make the connection process faster. As shown in the sample script in Listing \ref{lst:1}, in line 2 a IPv4 address is used in case of WLAN connection, while in line 3 a IPv6 address and the network interface alias (eth0) are specified, in case of LAN connection.

\subsection{Data streaming}
The script \textit{streaming.py} shows how to access live data. Specifically, the methods \textit{start\_streaming()} and \textit{stop\_streaming()} allow the developer to control the streaming mode of the eye-tracker (see also Listing \ref{lst:1}, lines 25 and 43). The \textit{TobiiGlassesController} deals with collecting gaze data in a Python dictionary for making them accessible using the method \textit{get\_data()} (see also Listing \ref{lst:1}, line 37).

\subsection{Managing projects, calibrations, participants and recordings}
The script \textit{calibrate\_and\_record.py} is a complete example of managing recordings. According to the API, a recording is associated to a single participant, it belongs to a specific project and it requires a successful calibration process. The example shows how to first create a project, a participant profile, how to complete a calibration and finally to control a recording. The script allows accessing the absolute path where the recording are stored in the SD card. The functionalities just mentioned are present in the Listing \ref{lst:1} as well. Differently, in order to retrieve information about already stored recordings, we implemented a set of methods for retrieving information of projects, recordings and segments, as shown in the examples \textit{01\_get\_projects.py}, \textit{02\_get\_recordings.py} and \textit{03\_get\_segments.py}.

\subsection{Live video scene and mapped gaze}
Live video streaming of the \textit{Scene Camera} is accessible through a RTSP (Real-Time Streaming Protocol) server running in the Recording Unit. The \textit{live\_scene.py} uses OpenCV \cite{opencv_library} for showing the on-line video from the scene camera, while the \textit{livescene\_and\_gaze.py} is an example of on-line mapping the gaze data with the video. In the other hand, the example \textit{video\_and\_gaze.py} implements the off-line mapping between video and gaze data. 

\subsection{Sending custom events}\label{sec:55}
During a recording is possible to send external triggers that indicate the occurrence of specific events. The API allows to send specific JSON objects containing information about these events. We implemented two mechanisms for sending custom events: the first one lets the user to have these information as logged events in the CSV-files (Listing \ref{lst:1}, lines 28 and 39), the second one adds the possibility to have custom events in Tobii Pro Lab (Listing \ref{lst:1}, lines 29 and 40) as well.

\vspace*{0.2in}
\begin{minipage}{0.95\linewidth}
\begin{lstlisting}[language=Python, label={lst:2}, caption=Python example about how exporting gaze data from a stored recording, frame=single]
from tobiiglasses.recordings import Recording
from tobiiglasses.filters.fixationsDT import FilterDT

rec = Recording(project_dir='data/', project_id='xejxnds', recording_id='k3l4jms')
rec.exportRawData(filename='rawdata.csv')
ff = FilterDT(dispersion_threshold=5, duration_threshold=100)
rec.exportFixations(fixation_filter=ff, filename='fdata.csv')
\end{lstlisting}
\end{minipage}

\subsection{Exporting data in CSV files}
Once a recording is successfully stored in the SD card of the \textit{Recording Unit}, the experimenter can proceed exporting the data using our suite. In the Listing \ref{lst:2} is shown how to export raw and fixations data from a stored recording in a CSV-file. The output CSV-file \textit{rawdata.csv} contains all the valid samples present in the livedata.json.gz, including gaze data, IMU data, and logged events. On the other hand, the \textit{fdata.csv} contains information about fixations points with relative duration, filtered by the I-DT filter. Additional exporting functions are shown in the example \textit{export\_data.py}.

\section{Conclusion}
A Python based open-source suite for using the Tobii Pro Glasses 2 has been presented in this paper. The availability of the source code in the open access repository and the possibility of installing the entire suite with a single Python package make our solution ready for use and especially suitable for the researcher's requirements. They can benefit from many examples and a full example experiment provided in the suite, aiming at facilitating the design of custom eye-tracking studies. In addition, the modularity of the software architecture and the use of a development platform for hosting the source code intend to ease future contributions from active developers, including bug fixes. The proposed solution is still under development, so it has great room for improvement in terms of robustness and in the number of implemented features. As future works, three main features are planned to be implemented. Firstly, the possibility to manage AOIs in the recordings. Secondly, the possibility to export eye-tracking metrics for fixations and AOIs. Thirdly, a set of visualization tools allowing to export heat maps and gaze plots.

\begin{acks}
This project has received funding from the European Research Council (ERC) under the European Union's Horizon 2020 research and innovation programme (grant awarded to AW, titled "InStance: Intentional Stance for Social Attunement". Grant agreement No: 715058).
\end{acks}

\bibliographystyle{ACM-Reference-Format}
\bibliography{main}
\end{document}